\documentclass[a4paper,twocolumn,accepted=2023-08-21]{quantumarticle}
\pdfoutput=1

\usepackage[utf8]{inputenc}
\usepackage[english]{babel}
\usepackage[a4paper, margin=1in]{geometry}
\usepackage{graphicx}
\usepackage{amsmath}
\usepackage{amsfonts}
\usepackage{physics}
\usepackage{bbold}
\usepackage{bm}
\usepackage{xcolor}
\usepackage{subcaption}
\usepackage{tikz}
\usetikzlibrary{matrix}
\usepackage{pgfplots}
\pgfplotsset{compat=1.16}
\pgfplotsset{scaled x ticks=false}
\definecolor{blue}{HTML}{0000FF}
\definecolor{orange}{HTML}{FFA500}
\definecolor{green}{HTML}{008000}
\definecolor{red}{HTML}{FF0000}
\usepackage{algorithm}
\usepackage{algpseudocode}

\usepackage[numbers, sort&compress]{natbib}
\usepackage{hyperref}
\usepackage{cleveref}
\crefformat{equation}{(#2#1#3)} 

\newcommand{\flip}{{\tt flip}}
\newcommand{\pflip}{p-{\tt flip}}

\begin{document}

\title{Local Probabilistic Decoding of a Quantum Code}
\author{T. R. Scruby}
\author{K. Nemoto}
\date{%
    \small{\textit{
    Okinawa Institute of Science and Technology, Okinawa, 904-0495, Japan\\%
    }}
    }

\maketitle

\begin{abstract}
    \flip{} is an extremely simple and maximally local classical decoder which has been used to great effect in certain classes of classical codes. When applied to quantum codes there exist constant-weight errors (such as half of a stabiliser) which are uncorrectable for this decoder, so previous studies have considered modified versions of \flip{}, sometimes in conjunction with other decoders. We argue that this may not always be necessary, and present numerical evidence for the existence of a threshold for \flip{} when applied to the looplike syndromes of a three-dimensional toric code on a cubic lattice. This result can be attributed to the fact that the lowest-weight uncorrectable errors for this decoder are closer (in terms of Hamming distance) to correctable errors than to other uncorrectable errors, and so they are likely to become correctable in future code cycles after transformation by additional noise. Introducing randomness into the decoder can allow it to correct these ``uncorrectable'' errors with finite probability, and for a decoding strategy that uses a combination of belief propagation and probabilistic \flip{} we observe a threshold of $\sim5.5\%$ under phenomenological noise. This is comparable to the best known threshold for this code ($\sim7.1\%$) which was achieved using belief propagation and ordered statistics decoding~[Higgott and Breuckmann, 2022], a strategy with a runtime of $O(n^3)$ as opposed to the $O(n)$ ($O(1)$ when parallelised) runtime of our local decoder. We expect that this strategy could be generalised to work well in other low-density parity check codes, and hope that these results will prompt investigation of other previously overlooked decoders.
\end{abstract}

\section{Introduction}
\label{section:intro}
An error correcting code is a method of encoding $k$ bits of logical information into $n$ physical bits in such a way that the redundancy of this encoding allows for the detection and correction of errors affecting these physical bits. Detection of errors is commonly achieved through the use of parity checks, which are rules stipulating that the overall parities of certain subsets of bits should be even. Configurations of bits satisfying all parity checks are called codewords of the code. Unsatisfied parity checks indicate the presence of errors, and given knowledge of whether or not each parity check is satisfied (this information is called a syndrome) we can attempt to calculate a correction. An algorithm which performs this calculation is called a decoder. The distance, $d$, of a code is defined as the minimum Hamming distance between codewords, and the best possible decoder for a given code will be capable of correcting any error affecting fewer than $d/2$ bits~\cite{mackay_information_nodate}.

In a quantum error correcting code the parity checks are replaced by operators called stabilisers, which are multi-qubit Pauli measurements with expected outcome $+1$ (in the absence of errors)~\cite{gottesman_stabilizer_1997}. In the class of quantum codes called Calderbank-Shor-Steane (CSS) codes~\cite{calderbank_good_1996,steane_multiple_1996} every stabiliser consists of only $X$ or only $Z$ measurements and the quantum decoding problem can be viewed as two separate classical decoding problems (one for $Z$ errors and one for $X$ errors). Classical decoders can (in principle) be applied to these problems directly, but we generally do not expect unmodified classical decoders to perform well in this setting because the syndromes of quantum codes are degenerate, meaning that there are multiple equivalent\footnote{Two errors $E_1$ and $E_2$ in a quantum code are equivalent if $E_1S = E_2$ where $S$ is a stabiliser of the code.} errors with the same syndrome~\cite{leverrier_quantum_2015,poulin_iterative_2008}. In particular, because stabilisers act trivially on the encoded information any error $E$ which is equivalent to half of a stabiliser ($S$) can be corrected equally well by either applying $E$ again or by applying the other half of the stabiliser ($ES$). The fact that both of these options have the same weight (they act non-trivially on the same number of qubits) as well as the same syndrome can make it difficult for classical decoders to choose between the two, so classical decoders are often modified or combined with other decoders when being applied to quantum codes~\cite{leverrier_quantum_2015,grospellier_numerical_2019,grospellier_combining_2021,poulin_iterative_2008,li_numerical_2020,panteleev_degenerate_2021,roffe_decoding_2020,quintavalle_single-shot_2021,higgott_improved_2022}. 

In this work we examine the necessity of these modifications by simulating the performance of two classical decoders (\flip{}~\cite{sipser_expander_1996} and BP~\cite{mackay_information_nodate}) when applied to the three-dimensional toric code (3DTC). Both of these decoders suffer from the problem described above (they are unable to correct errors equivalent to half of a stabiliser) and \flip{} suffers from the additional problem that it cannot correct any membrane of errors whose boundary is sufficiently smooth. Despite this, in both cases we observe thresholds competitive with those obtained by some of the best known decoders for this code. This result can be understood using the same intuition that helps explain the validity of single-shot decoding~\cite{bombin_single-shot_2015,quintavalle_single-shot_2021} and self-correcting quantum memories~\cite{dennis_topological_2002}\footnote{In fact, one of the decoders we investigate is almost identical to the recovery procedure proposed in~\cite{dennis_topological_2002} for the self-correcting four-dimensional toric code}; our recovery procedure does not need to remove all errors, it only needs to remove sufficiently many that it achieves a suitably low-error steady state in conjunction with environmental noise. At the end of the computation the qubits of the code will be measured out and a more sophisticated decoder (capable of correcting all constant-size errors) can be used to process this information before declaring the outcome of the logical measurement. As long as the average number of errors in the code at steady-state is below the amount that can be corrected by this more sophisticated decoder we expect that we can reliably read out the encoded logical information.

The rest of this manuscript is arranged as follows: in \cref{section:3D} we provide a brief overview of the 3D toric code, and in \cref{section:flip} we examine the performance of \flip{} (and a non-deterministic variant we call \pflip{}) when used to correct errors in this code. In \cref{section:BP} we perform a similar examination of the belief-propagation (BP) decoder and a hybrid BP+\pflip{} decoder. Finally, we discuss possible improvements and outlooks in \cref{section:discussion}. 

All code used in obtaining the results presented in this paper can be found at~\cite{source_code}, and additional discussion of simulation methods can be found in \cref{appendix:sim}

\section{The 3D Toric Code}
\label{section:3D}
The familiar 2D toric code (2DTC) is obtained by taking a square lattice with periodic boundaries and placing a qubit on each edge, an $X$ stabiliser on each vertex (so that it is supported on the four edges that meet at this vertex) and a $Z$ stabiliser on each face~\cite{kitaev_fault-tolerant_2003}. The 3DTC is a straightforward generalisation obtained by placing qubits on the edges of a cubic lattice, with $X$ ($Z$) stabilisers once again associated with vertices (faces)~\cite{dennis_topological_2002}.

\begin{figure}
    \centering
    \begin{subfigure}{.2\textwidth}
    \includegraphics[width=\textwidth]{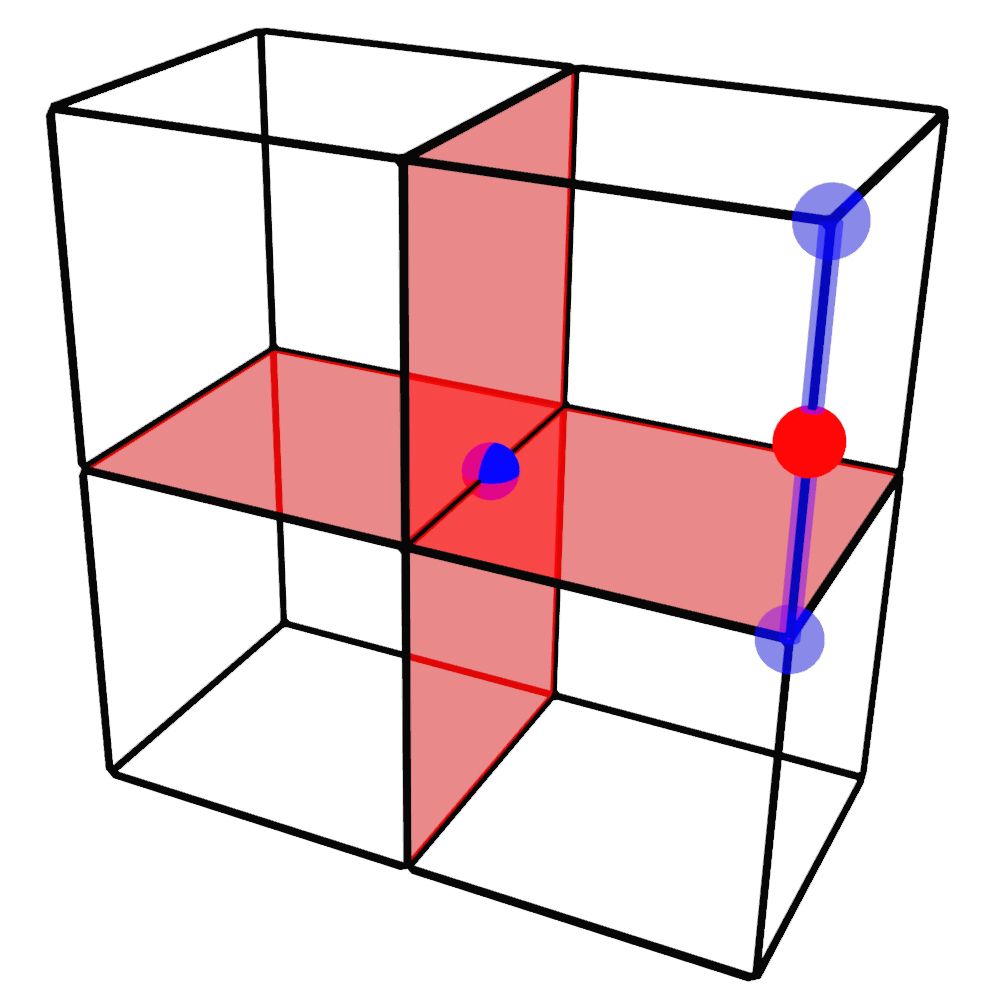}
    \subcaption{}
    \label{subfig:3dtc_a}
    \end{subfigure}
    ~
    \begin{subfigure}{.2\textwidth}
    \includegraphics[width=\textwidth]{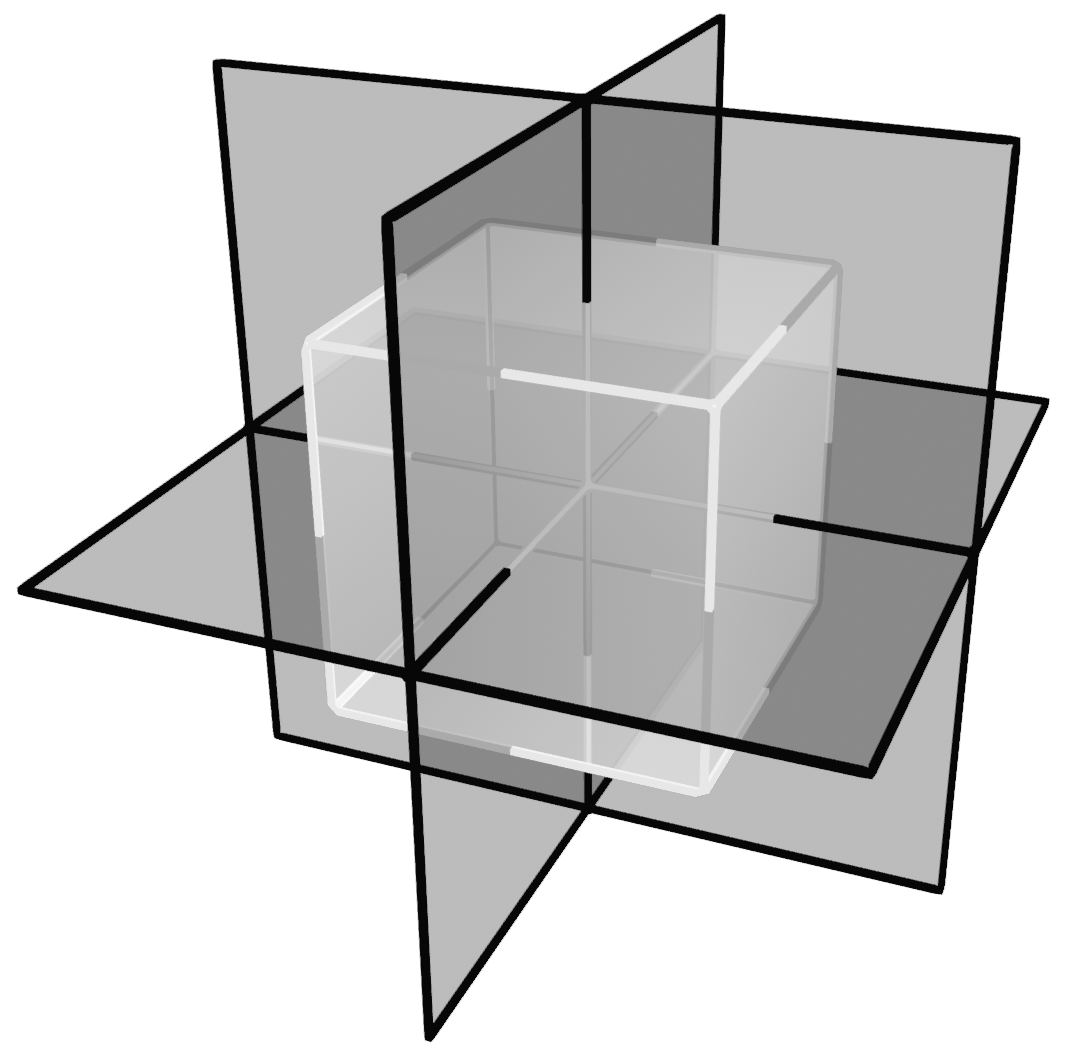}
    \subcaption{}
    \label{subfig:3dtc_b}
    \end{subfigure}
    \caption{(a) Four cells of the 3D toric code. Qubits, $X$ stabilisers and $Z$ stabilisers are assigned to edges, vertices and faces respectively. Single-qubit $X$ and $Z$ errors (solid blue and red) and their syndromes (transparent red and blue) are also shown. (b) Correspondence between the vertex neighbourhood of a 3D cubic lattice (dark faces and edges) and its dual (light faces and edges).}
    \label{fig:3dtc}
\end{figure}

Unlike in the 2DTC (where each $X$ and $Z$ stabiliser is supported on four qubits and each qubit is part of two $X$ and two $Z$ stabilisers), the 3DTC has an asymmetry between $X$ and $Z$ stabilisers. Namely, each $X$ stabiliser is supported on six qubits and each qubit is part of two $X$ stabilisers, whereas each $Z$ stabiliser is supported on four qubits and each qubit is part of four $Z$ stabilisers. This last detail is of particular significance as it means that $X$ errors in the code produce syndromes which are looplike, rather than pointlike, and thus can be corrected using a \textit{single-shot} decoder~\cite{bombin_single-shot_2015,quintavalle_single-shot_2021}. Examples of errors and their syndromes are shown in \cref{subfig:3dtc_a}. 

If we are concerned only with $X$ errors (as will be the case in what follows) it is more convenient to describe the code using the dual lattice, in which cells are exchanged with vertices and edges are exchanged with faces. An example of this mapping is shown in \cref{subfig:3dtc_b}. In this lattice qubits are associated with faces while $Z$ stabilisers are associated with edges, meaning that an $X$ error on a qubit/face will be detected by the $Z$ stabilisers on the four edges of this face. 

There is also an asymmetry between the logical operators of this code: the logical $X$ operators are membranes which wrap around the 3-torus in two directions while the logical $Z$ operators are strings normal to these membranes which wrap around the 3-torus in the third direction. There are three possible planes in which these membranes can lie and thus there are three encoded qubits. In what follows we will consider torii obtained by imposing periodic boundary conditions on an $L \times L \times L$ cubic lattice. This results in a minimum weight of $L^2$ for logical $X$ operators and a minimum weight of $L$ for logical $Z$ operators, giving codes with parameters $[\![3L^3,3,L]\!]$.

\section{\flip{} and \pflip{}}
\label{section:flip}
\subsection{\flip{}}

\flip{} is an extremely simple and maximally local decoder first proposed by Sipser and Spielman~\cite{sipser_expander_1996} for use with classical expander codes. The decoder works in the following way: define $\mathrm{SAT}(b)$ to be the set of satisfied checks containing bit $b$ and $\mathrm{UNSAT}(b)$ to be the set of unsatisfied checks containing bit $b$, then do
\newline

\begin{algorithmic}
    \While{$\exists ~ b \in \textbf{bits } \textrm{s.t. } |\mathrm{SAT}(b)| < |\mathrm{UNSAT}(b)|$}
        \State flip $b$
    \EndWhile
\end{algorithmic}

In other words, for each bit of the code we count the number of unsatisfied and satisfied checks to which that bit belongs, flip the bit if there are more of the former than the latter, and repeat until there are no such bits. The sequential version of this decoder runs in time linear in the total number of bits, while a parallelised version runs in constant time. 

\begin{figure}
    \centering
        \begin{tikzpicture}
            \begin{axis}[
                xlabel={$p$},
                ylabel={$p_{\mathrm{fail}}(p,L)$},
                ymode=log,
                width=7cm,
                height=6.5cm,
                ymin=0.00011,
                ymax=1.5,
                xmin=0.02532,
                xtick={0.0255,0.026,0.0265,0.027},
                minor xtick={0.0254,0.0256,0.0258,0.0262,0.0264,0.0266,0.0268},
                xticklabels={$2.55\%$,$2.6\%$,$2.65\%$,$2.7\%$},
                legend style={
                    at={(0.95,0.05)},
                    anchor=south east
                }
            ]
            \addplot[
                color=blue,
                mark=*,
                mark size=1.5,
                error bars/.cd,
                y dir=both,
                y explicit
            ] table [x=p, y=pfail, y error=err, col sep=comma] {data/flip_1-2/processed14.csv};
            \addplot[
                color=orange,
                mark=triangle*,
                mark size=2,
                error bars/.cd,
                y dir=both,
                y explicit
            ] table [x=p, y=pfail, y error=err, col sep=comma] {data/flip_1-2/processed18.csv};
            \addplot[
                color=green,
                mark=diamond*,
                mark size=2,
                error bars/.cd,
                y dir=both,
                y explicit
            ] table [x=p, y=pfail, y error=err, col sep=comma] {data/flip_1-2/processed24.csv};
            \addplot[
                color=red,
                mark=square*,
                mark size=1.2,
                error bars/.cd,
                y dir=both,
                y explicit
            ] table [x=p, y=pfail, y error=err, col sep=comma] {data/flip_1-2/processed32.csv};
            \legend{$L=14$, $L=18$, $L=24$, $L=32$}
            \end{axis}
        \end{tikzpicture}
    \caption{Performance of the parallelised \flip{} decoder in the 3DTC for physical $X$ error rate $p$ equal to $Z$ stabiliser measurement error rate $q$. $p_{fail}$ is the logical $X$ error rate calculated via a majority vote on the measurement outcomes of a complete set of disjoint logical $Z$ operators after 1000 code cycles. \flip{} is applied simultaneously to all qubits of the code once per code cycle.}
    \label{fig:flip_1-2}
\end{figure}
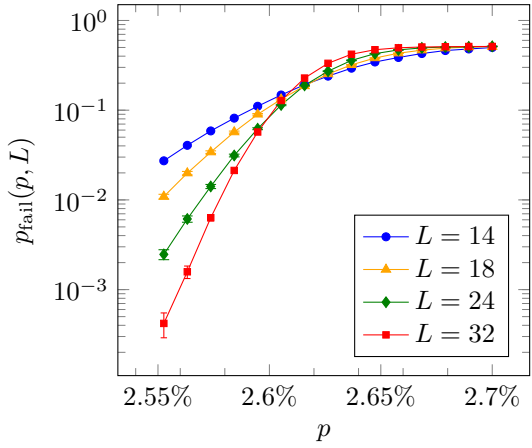

For \flip{} to perform well we require that each bit is involved in a relatively large number of checks. For an example of a case where \flip{} does not perform well we can consider a cyclic repetition code. Here each bit is part of two checks, meaning that if an error occurs on two bits of the same check then each of these bits will be part of one satisfied and one unsatisfied check and \flip{} will not be able to correct this error.

If we consider the syndromes produced by $Z$ errors in the 3DTC we see that they have exactly the same problem. Each qubit is part of two $X$ stabilisers and thus strings of $Z$ errors will only be detected at their endpoints, meaning that for any string of length $> 2$ there are no qubits which are part of more unsatisfied than satisfied $X$ checks (unless the ends of the string are very close together). On the other hand, every qubit is part of four $Z$ stabilisers and the syndromes produced by $X$ errors are looplike, meaning the weight of the syndrome (i.e. number of unsatisfied checks) scales with the weight of the error. Intuitively we might therefore expect \flip{} to be effective in this case since corrections which decrease the weight of a given syndrome will also decrease the weight of the corresponding error. 

\begin{figure}
    \centering
    \begin{subfigure}{0.2\textwidth}
    \begin{tikzpicture}   
    \draw[black] (0,0) -- (1,1) -- (3,1) -- (2,0) -- cycle;                              
    \draw[black] (0,0) -- (0,2) -- (1,3) -- (1,1) -- cycle;                              
    \draw[black] (0,0) -- (0,2) -- (2,2) -- (2,0) -- cycle;                              
    \draw[black,fill=blue,opacity=0.6] (1,1) -- (1,3) -- (3,3) -- (3,1) -- cycle;        
    \draw[red,line width=0.15cm,rounded corners,line cap=round] (3,1) -- (1,1) -- (1,3);
    \draw[black,fill=blue,opacity=0.6] (0,2) -- (1,3) -- (3,3) -- (2,2) -- cycle;        
    \draw[black,fill=blue,opacity=0.6] (2,0) -- (2,2) -- (3,3) -- (3,1) -- cycle;        
    \draw[red,line width=0.15cm,rounded corners,line cap=round] (1,3) -- (0,2) -- (2,2) -- (2,0) -- (3,1);
    \end{tikzpicture}
    \subcaption{}
    \label{subfig:uncorrec_a}
    \end{subfigure}
    ~
    \begin{subfigure}{0.2\textwidth}
    \begin{tikzpicture}
    \draw[black] (1.5,0) -- (1.5,3);
    \draw[black] (0,1.5) -- (3,1.5);
    \draw[black,fill=blue,opacity=0.6] (0,0) -- (0,3) -- (3,3) -- (3,0) -- cycle;
    \draw[red,line width=0.15cm,rounded corners] (0,0) -- (0,3) -- (3,3) -- (3,0) -- cycle;
    \end{tikzpicture}
    \subcaption{}
    \label{subfig:uncorrec_b}
    \end{subfigure}
    \caption{Two examples of low-weight errors which are uncorrectable for \flip{}. Blue faces are qubits subject to $X$ errors and red edges are unsatisfied $Z$ checks. (a) is a three-qubit $X$ error on three of the six qubits of an $X$ stabiliser. (b) is a four-qubit error with all four qubits lying in a plane. In both cases each qubit is part of two satisfied and two unsatisfied $Z$ checks and so \flip{} will be unable to find a correction.}
    \label{fig:uncorrec}
\end{figure}
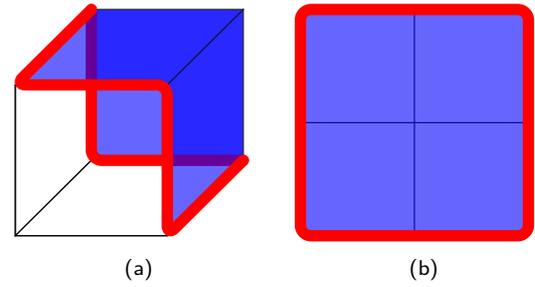

\begin{figure*}
    \centering
    \begin{subfigure}{.24\textwidth}
    \centering
    \begin{tikzpicture}
    \draw[black] (0,0) -- (1,1) -- (3,1) -- (2,0) -- cycle;                              
    \draw[black] (0,0) -- (0,2) -- (2,2) -- (2,0) -- cycle;                              
    \draw[black,fill=blue,opacity=0.6] (1,1) -- (1,3) -- (3,3) -- (3,1) -- cycle;        
    \draw[red,line width=0.15cm,rounded corners,line cap=round] (1,1) -- (3,1);
    \draw[black,fill=blue,opacity=0.6] (0,0) -- (0,2) -- (1,3) -- (1,1) -- cycle;        
    \draw[black,fill=blue,opacity=0.6] (0,2) -- (1,3) -- (3,3) -- (2,2) -- cycle;        
    \draw[black,fill=blue,opacity=0.6] (2,0) -- (2,2) -- (3,3) -- (3,1) -- cycle;        
    \draw[red,line width=0.15cm,rounded corners,line cap=round] (3,1) -- (2,0) -- (2,2) -- (0,2) -- (0,0) -- (1,1);
    \end{tikzpicture}
    \subcaption{}
    \label{subfig:newerrors_a}
    \end{subfigure}
    ~
    \begin{subfigure}{.36\textwidth}
    \centering
    \begin{tikzpicture}
    \draw[black] (0,0) -- (1,1) -- (3,1) -- (2,0) -- cycle;                              
    \draw[black] (0,0) -- (0,2) -- (1,3) -- (1,1) -- cycle;                              
    \draw[black] (0,0) -- (0,2) -- (2,2) -- (2,0) -- cycle;                              
    \draw[black,fill=blue,opacity=0.6] (1,1) -- (1,3) -- (3,3) -- (3,1) -- cycle;        
    \draw[red,line width=0.15cm,rounded corners,line cap=round] (3,1) -- (1,1) -- (1,3);
    \draw[black,fill=blue,opacity=0.6] (0,2) -- (1,3) -- (3,3) -- (2,2) -- cycle;        
    \draw[black,fill=blue,opacity=0.6] (2,0) -- (2,2) -- (3,3) -- (3,1) -- cycle;        
    \draw[black,fill=blue,opacity=0.6] (0,2) -- (-2,2) -- (-1,3) -- (1,3) -- cycle;      
    \draw[red,line width=0.15cm,rounded corners,line cap=round] (1,3) -- (-1,3) -- (-2,2) -- (2,2) -- (2,0) -- (3,1);
    \end{tikzpicture}
    \subcaption{}
    \label{subfig:newerrors_b}
    \end{subfigure}
    ~
    \begin{subfigure}{.36\textwidth}
    \centering
    \begin{tikzpicture}
    \draw[black] (0,0) -- (1,1) -- (3,1) -- (2,0) -- cycle;                              
    \draw[black] (0,0) -- (0,2) -- (1,3) -- (1,1) -- cycle;                              
    \draw[black] (0,0) -- (0,2) -- (2,2) -- (2,0) -- cycle;                              
    \draw[black,fill=blue,opacity=0.6] (1,1) -- (1,3) -- (3,3) -- (3,1) -- cycle;        
    \draw[black,fill=blue,opacity=0.6] (1,1) -- (-1,1) -- (-1,3) -- (1,3) -- cycle;      
    \draw[red,line width=0.15cm,rounded corners,line cap=round] (3,1) -- (-1,1) -- (-1,3);
    \draw[black,fill=blue,opacity=0.6] (0,2) -- (1,3) -- (3,3) -- (2,2) -- cycle;        
    \draw[black,fill=blue,opacity=0.6] (2,0) -- (2,2) -- (3,3) -- (3,1) -- cycle;        
    \draw[black,fill=blue,opacity=0.6] (0,2) -- (-2,2) -- (-1,3) -- (1,3) -- cycle;      
    \draw[red,line width=0.15cm,rounded corners,line cap=round] (-1,3) -- (-2,2) -- (2,2) -- (2,0) -- (3,1);
    \end{tikzpicture}
    \subcaption{}
    \label{subfig:newerrors_c}
    \end{subfigure}
    ~
    \begin{subfigure}{.24\textwidth}
    \centering
    \begin{tikzpicture}
    \draw[black,fill=blue,opacity=0.6] (0,0) -- (0,3) -- (1.5,3) -- (1.5,1.5) -- (3,1.5) -- (3,0) -- cycle;
    \draw[black] (0,1.5) -- (1.5,1.5) -- (1.5,0);
    \draw[red,line width=0.15cm,rounded corners] (0,0) -- (0,3) -- (1.5,3) -- (1.5,1.5) -- (3,1.5) -- (3,0) -- cycle;
    \end{tikzpicture}
    \subcaption{}
    \label{subfig:newerrors_d}
    \end{subfigure}
    ~
    \begin{subfigure}{.36\textwidth}
    \centering
    \begin{tikzpicture}
    \draw[black,fill=blue,opacity=0.6] (0,0) -- (4.5,0) -- (4.5,1.5) -- (3,1.5) -- (3,3) -- (0,3) -- cycle;
    \draw[black] (1.5,0) -- (1.5,3);
    \draw[black] (0,1.5) -- (3,1.5) -- (3,0);
    \draw[red,line width=0.15cm,rounded corners] (0,0) -- (4.5,0) -- (4.5,1.5) -- (3,1.5) -- (3,3) -- (0,3) -- cycle;
    \end{tikzpicture}
    \subcaption{}
    \label{subfig:newerrors_e}
    \end{subfigure}
    ~
    \begin{subfigure}{.36\textwidth}
    \centering
    \begin{tikzpicture}
    \draw[black,fill=blue,opacity=0.6] (0,0) -- (4.5,0) -- (4.5,3) -- (0,3) -- cycle;
    \draw[black] (1.5,0) -- (1.5,3);
    \draw[black] (3,0) -- (3,3);
    \draw[black] (0,1.5) -- (4.5,1.5);
    \draw[red,line width=0.15cm,rounded corners] (0,0) -- (4.5,0) -- (4.5,3) -- (0,3) -- cycle;
    \end{tikzpicture}
    \subcaption{}
    \label{subfig:newerrors_f}
    \end{subfigure}
    \caption{(a)-(c) Some possible additional errors modifying the uncorrectable error shown in \cref{subfig:uncorrec_a}. In (a) an additional error on a fourth face of the cube makes the error correctable (an error on any of the cube's six faces has the same effect). In (b) an error occurs on a qubit next to the cube. \flip{} will correct this additional error and restore the original three-qubit error. In (c) a pair of errors on qubits next to the cube grow the original error into a five-qubit uncorrectable error. (d) - (f) show equivalent situations for the four-qubit uncorrectable error shown in \cref{subfig:uncorrec_b}}
    \label{fig:newerrors}
\end{figure*}
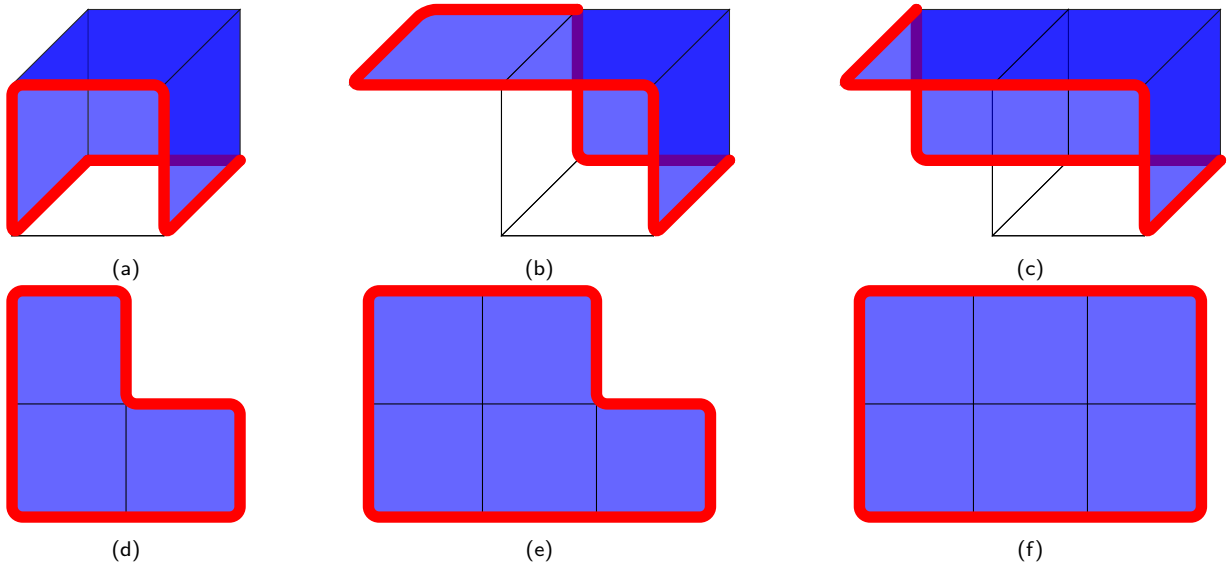

\Cref{fig:flip_1-2} shows the simulated performance of a parallelised version of \flip{} when used to decode $X$ errors in the 3DTC. In these simulations we evaluate the flip rule only once per qubit per code cycle rather than running the decoder until it runs out of qubits to flip (``flipping'' a qubit means applying a single-qubit $X$ correction to that qubit). Presence/absence of a logical $X$ error at the end of the simulation is determined via majority vote on the measurement outcomes of a complete disjoint set of logical $Z$ operators\footnote{The results shown here are for only one of the three logical qubits encoded by the 3DTC, but as the code, error model and decoder are symmetric along all three axes the logical error rates for the other two encoded qubits will be identical. Correlations between the error rates of these encoded qubits are discussed in \cref{appendix:sim}}. We observe that these results agree with our previously stated intuition. For a phenomenological noise model with $X$ error probability per qubit ($p$) equal to measurement error probability per $Z$ stabiliser ($q$) we observe a threshold of $p_{\mathrm{th}} \sim 2.6\%$ after 1000 code cycles. This is comparable with the thresholds obtained for the BP+OSD decoder in~\cite{quintavalle_single-shot_2021} (although higher thresholds for a modified version of this decoder were recently obtained in~\cite{higgott_improved_2022}). Some readers might find this result surprising since, as mentioned previously, we do not generally expect classical decoders to perform well when naively applied to quantum codes and in this specific case we can easily find examples of low-weight $X$ errors which cannot be corrected by \flip{} (two examples are shown in \cref{fig:uncorrec}). Why, then, does \flip{} seem to perform so well?

The answer is that all of the lowest weight (and therefore most likely) uncorrectable errors are closer (in terms of Hamming weight) to correctable errors than they are to other uncorrectable errors. As a result, additional errors occurring in the vicinity of such a low-weight uncorrectable error will transform it into a correctable error with high probability. Examples are shown in \cref{fig:newerrors}. Here we see how, for the low-weight uncorrectable errors shown in \cref{fig:uncorrec}, additional single-qubit $X$ errors can make these errors correctable but cannot transform them into larger uncorrectable errors. The behaviour of the code under the combined action of \flip{} + environmental noise is then analagous to a physical system relaxing towards its ground state at finite temperature, with small thermal fluctuations allowing it to escape from shallow local minima. 

We comment at this point that despite the impressive performance observed for \flip{} in this setting we do not expect to observe a sustainable threshold (as discussed in~\cite{brown_fault-tolerant_2016}, for example) for this decoder. This is due to the fact that even though the transformation of low-weight uncorrectable errors to higher-weight ones is relatively unlikely, we do expect it to occur over a long enough timescale. Larger uncorrectable errors require more additional errors to make them correctable, and so we expect that regions of uncorrectable errors will slowly build up in the code over many code cycles. However, the results in \cref{fig:flip_1-2} indicate that this buildup is very slow and so a sustainable threshold could be recovered by combining \flip{} with another decoder capable of correcting all errors below a certain weight. By running this more accurate (but also more complex) decoder every $m$ code cycles we would remove the buildup of uncorrectable errors. $m$ can be chosen to be very large without significantly compromising performance (e.g. $m=1000$ should give at least the threshold of $\sim 2.6\%$ observed in \cref{fig:flip_1-2} provided the more complex decoder has an equivalent or higher threshold) and so this would represent a significant decrease in decoding time/complexity relative to using the more complex decoder at every code cycle. 

\subsection{\pflip{}}

Having seen the power of random (qu)bit flips in transforming uncorrectable errors into correctable ones we now consider the incorporation of these random flips into the decoder itself. Consider the following, modified version of \flip{} which we will refer to as probabilistic \flip{} or \pflip{}.

\begin{algorithmic}
    \For{$b \in \textbf{bits }$}
        \If{$|\mathrm{SAT}(b)| < |\mathrm{UNSAT}(b)|$}
            \State flip $b$
        \ElsIf{$|\mathrm{SAT}(b)| = |\mathrm{UNSAT}(b)|$}
            \State $x \gets$ random number in (0,1)
            \If{x $<$ 0.5}
                \State flip $b$
            \EndIf
        \EndIf
    \EndFor
\end{algorithmic}

This decoder is (very nearly\footnote{The only difference is that this decoder acts on all qubits at once, whereas the decoder of~\cite{dennis_topological_2002} acts on some subset of the qubits chosen such that no two qubits in the set are part of the same $Z$ stabiliser}) identical to the local recovery procedure for the four-dimensional toric code proposed in~\cite{dennis_topological_2002}. The decoder acts identically to \flip{} on bits where $\mathrm{SAT}(b)$ and $\mathrm{UNSAT}(b)$ are different sizes, but when they are the same size it flips that bit with probability $p=0.5$. Consider the action of \pflip{} on the two uncorrectable errors from \cref{fig:uncorrec}. For \cref{subfig:uncorrec_a} each qubit in the cube is part of two satisfied and two unsatisfied $Z$ checks and so any of them might be flipped by \pflip{}. As we saw in \cref{fig:newerrors}, this would make the error correctable for \flip{} (and so also for the deterministic part of \pflip{}). Something very similar is true for \cref{subfig:uncorrec_b} as every qubit in the error's support is part of two satisfied and two unsatisfied $Z$ checks and flipping any of them makes the error correctable for \flip{}. Notice also that, in each case, these are the \textit{only} qubits which are part of an equal number of satisfied and unsatisfied $Z$ checks. Thus \pflip{} can never increase the size of these errors while leaving them uncorrectable. We say that these errors are \textit{probabilistically correctable} for \pflip{} while being uncorrectable for \flip{}.

\pflip{} is not a strict improvement over \flip{}, however. Consider, for example, a two-qubit error where the errors are on opposite faces of one of the $X$ stabiliser cubes of the code. \flip{} is guaranteed to correct this error with only one application of the decoding rule per qubit. \pflip{}, on the other hand, might flip any of the four error-free faces of the cube (as they are each part of two satisfied and two unsatisfied $Z$ checks) if applied to these faces before being applied to the other two. This could produce a weight-three error which would require an additional cycle to correct. We therefore conclude that \flip{} is better than \pflip{} at correcting determinisitically correctable errors and that \pflip{} should only be called once \flip{} gets stuck. 

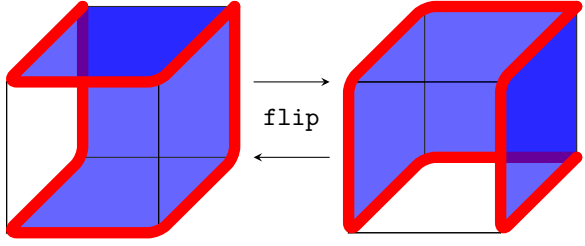
\begin{figure}
    \centering
    \begin{tikzpicture}
    \draw[black,fill=blue,opacity=0.6] (0,0) -- (1,1) -- (3,1) -- (2,0) -- cycle;        
    \draw[black] (0,0) -- (0,2) -- (1,3) -- (1,1) -- cycle;                              
    \draw[black] (0,0) -- (0,2) -- (2,2) -- (2,0) -- cycle;                              
    \draw[black,fill=blue,opacity=0.6] (1,1) -- (1,3) -- (3,3) -- (3,1) -- cycle;        
    \draw[black] (2,0) -- (2,2) -- (3,3) -- (3,1) -- cycle;                              
    \draw[red,line width=0.15cm,rounded corners,line cap=round] (1,3) -- (1,1) -- (0,0) -- (2,0) -- (3,1) -- (3,3);
    \draw[black,fill=blue,opacity=0.6] (0,2) -- (1,3) -- (3,3) -- (2,2) -- cycle;        
    \draw[red,line width=0.15cm,rounded corners,line cap=round] (1,3) -- (0,2) -- (2,2) -- (3,3);
    
    \draw[-stealth] (3.25,2) -- (4.25,2);
    \node at (3.75,1.5) {\flip{}};
    \draw[stealth-] (3.25,1) -- (4.25,1);
    
    \draw[black] (4.5,0) -- (5.5,1) -- (7.5,1) -- (6.5,0) -- cycle;                              
    \draw[black,fill=blue,opacity=0.6] (4.5,0) -- (4.5,2) -- (5.5,3) -- (5.5,1) -- cycle;        
    \draw[black] (4.5,0) -- (4.5,2) -- (6.5,2) -- (6.5,0) -- cycle;                              
    \draw[black,fill=blue,opacity=0.6] (5.5,1) -- (5.5,3) -- (7.5,3) -- (7.5,1) -- cycle;        
    \draw[black] (4.5,2) -- (5.5,3) -- (7.5,3) -- (6.5,2) -- cycle;                              
    \draw[red,line width=0.15cm,rounded corners,line cap=round] (7.5,1) -- (5.5,1) -- (4.5,0) -- (4.5,2) -- (5.5,3) -- (7.5,3);
    \draw[black,fill=blue,opacity=0.6] (6.5,0) -- (6.5,2) -- (7.5,3) -- (7.5,1) -- cycle;        
    \draw[red,line width=0.15cm,rounded corners,line cap=round] (7.5,3) -- (6.5,2) -- (6.5,0) -- (7.5,1);
    \end{tikzpicture}
    \caption{A pair three-qubit errors which are exchanged by the action of the parallelised version of \flip{}}
    \label{fig:uncorrec_dynamic}
\end{figure}

A distinction should also be made between the sequential and parallel versions of \flip{}/\pflip{}. Since the parallelisability of these decoders (resulting in decoding time constant in the code size) is one of their main strengths we will consider only the parallel case. In this case there is one additional type of weight-3 uncorrectable error that we need to be concerned with, which we show in \cref{fig:uncorrec_dynamic}. This error is correctable for the sequential version of \flip{} but not for the parallel version since there are four qubits (two with errors and two without) which each are part of one satisfied and three unsatisfied checks, and flipping all four of these qubits at once results in a rotated version of the original error. In addition to simply being uncorrectable these errors also prevent us from detecting that \flip{} is stuck, since their existence in the code means that \flip{} will never stop finding bits to flip. Thus, in a parallel implementation of \flip{} + \pflip{} we must apply \pflip{} on some fixed schedule rather than waiting for \flip{} to terminate.

To find an optimal schedule we simulate the performance of \flip{} + \pflip{} while varying the number of applications of the decoding rule per qubit per code cycle ($n_A$) and the (inverse) frequency with which we apply \pflip{} relative to \flip{} ($\lambda_p$). For example, the parameters $n_A = 5$ and $\lambda_p = 3$ describe the following schedule:

\begin{table}[]
    \centering
    \begin{tikzpicture}
    \matrix (T) [matrix of nodes,column sep=-\pgflinewidth,row sep=-\pgflinewidth,nodes={draw,minimum height=1cm}]
    {
      |[draw=none]| 9 & 4.343\% & 4.652\% & 4.753\% & 4.652\% & 4.654\% \\
      |[draw=none]| 7 & 4.212\% & 4.629\% & 4.615\% & 4.620\% & 4.558\% \\
      |[draw=none]| 5 & 3.998\% & 4.445\% & 4.475\% & 4.274\% & |[fill=gray]| 4.082\% \\
      |[draw=none]| 3 & 3.604\% & 3.752\% & |[fill=gray]| 3.943\% &   &   \\
      |[draw=none]| 1 & |[fill=gray]| 2.611\% &   &   &   &   \\
        & |[draw=none]| 2 & |[draw=none]| 3 & |[draw=none]| 4 & |[draw=none]| 5 & |[draw=none]| 6 \\
    };
    \node[rotate=90,anchor=south] at (T.west) {\hspace{1cm}Decoding rule applications $n_A$};
    \node[anchor=north] at (T.south) {\pflip{} inverse frequency $\lambda_p$};
    \end{tikzpicture}
    \caption
    [pflip thresholds]
    {Thresholds after 1000 code cycles for \flip{} + \pflip{} under phenomenological noise (with $p=q$) in the 3DTC for various values of the parameters $n_A$ and $\lambda_p$. Uncertainties on all values are $0.001\%$\footnotemark{} and were calculated by SciPy's curve\_fit function. Shaded cells are those where $\lambda_p > n_A$ (where the decoding strategy consists of $n_A$ rounds of \flip{}\footnotemark{}). We do not observe thresholds for the case of $\lambda_p = 1$ (where the decoding strategy consists of $n_A$ rounds of \pflip{}).} 
    \label{table:thresholds}
\end{table}

\addtocounter{footnote}{-1}
\footnotetext{More precisely, all uncertainties are in the range $[0.0005\%, 0.0015\%)$.}
\addtocounter{footnote}{1}
\footnotetext{Every $\lambda_p^{\textrm{th}}$ application of the decoding rule within a code cycle uses \pflip{} instead of \flip{}. As we only count within code cycles and not between them any choice of $n_A$ and $\lambda_p$ such that $\lambda_p > n_A$ refers to a decoding strategy that uses only $n_A$ applications of \flip{} per qubit per code cycle and does not use \pflip{}.}

\begin{itemize}
    \item Measure $Z$ stabilisers
    \item \flip{}
    \item \flip{}
    \item \pflip{}
    \item \flip{}
    \item \flip{}
    \item Apply correction
\end{itemize}

Calculated thresholds for a range of $n_A$ and $\lambda_p$ are displayed in \cref{table:thresholds}. These thresholds were calculated by fitting data from our simulations to the ansatz $p_\textrm{fail}(x) = a_0 + a_1x + a_2x^2$ with $x=(p-p_\mathrm{th})L^{1/\nu}$~\cite{wang2003}. We observe that for $n_A = 1,3$ use of \pflip{} gives no improvement in performance relative to using only \flip{}. For $n_A \geq 5$ we do observe some performance improvement, with the highest thresholds of $\sim 4.5\%$ achieved for $\lambda_p = 3,4$. 

\section{BP and p-BP}
\label{section:BP}
Another decoder which has been successfully modified to work with quantum codes (despite being unable to correct half-stabiliser errors) is the belief propagation (BP) decoder~\cite{mackay_information_nodate}. The highest known threshold for the 3DTC was obtained using a combination of BP and ordered statistics decoding (OSD)~\cite{higgott_improved_2022}, and BP+OSD has also been shown to work well when applied to hypergraph and lifted product codes~\cite{roffe_decoding_2020,panteleev_degenerate_2021,roffe_bias-tailored_2022}. Simulations of the performance of BP without OSD in the 3DTC were also performed in~\cite{higgott_improved_2022}, where it was concluded that the decoder was unable to achieve a threshold. However, in that case BP was used for both at-runtime decoding and postprocessing/readout of logical information. In contrast, Grospellier et al.~\cite{grospellier_combining_2021} performed simulations of a hypergraph product code where BP was used only for at-runtime decoding of noisy syndrome data and a different decoding strategy was used to check for logical errors (we took the same approach with \flip{} in the previous section). In this case it was concluded that BP did possess a threshold and so it is natural to ask if the same can be achieved in the 3DTC. 

Before answering this question we provide a brief description of the BP decoder. We then present results demonstrating the existence of a threshold for pure BP decoding with readout performed via majority vote on logical $Z$ measurement outcomes and finally we discuss a hybrid decoding strategy that uses both BP and \pflip{}. 

\subsection{BP}

BP is a message-passing algorithm for calculating marginal probabilities of a distribution described by a factor graph~\cite{mackay_information_nodate}. Marginals are calculated exactly when the factor graph is a tree, but a good approximation to the exact answer can be obtained even if the graph contains loops. When BP is used as a decoding algorithm for a classical code (or for decoding only one type of error in a quantum CSS code) this graph will be the Tanner graph of the code, with one variable node for each bit and one factor node for each check. At every iteration of the algorithm each factor node passes messages to each adjacent variable node based on the messages the factor node received from all other adjacent variable nodes. Each variable node then passes messages to each adjacent factor node based on the messages received from all other factor nodes. In the general version of the algorithm one message must be sent for each possible state of the sending/receiving variable node, but in our case there are only two possible values for each variable (0 and 1) and there then exists a simpler version of the algorithm that requires only one message per edge of the graph at each step~\cite{chen_reduced-complexity_2005}. 

We begin by defining the log-likelihood ratio (LLR)

\begin{equation}
    l_i = \mathrm{log}\left(\frac{P(v_i = 0)}{P(v_i = 1)}\right)
\end{equation}

\noindent where $P(v_i = 0)$ is the probability that bit $i$ is error-free while $P(v_i = 1)$ is the probability that this bit has been flipped. We use $l_i^0$ to denote the LLR calculated using an initial guess about these probabilities based on an assumed error channel, e.g. for the binary symmetric channel (each bit flipped with probability $p$) $l_i^0=\mathrm{log}((1-p)/p)$ for each bit. These LLRs will be the original variable-to-factor messages ($M_{v_i \rightarrow f_j}$) used to initialise the algorithm. The factor-to-variable messages are then defined as

\begin{equation}
    M_{f_j \rightarrow v_i} = (-1)^{s_j}2\mathrm{tanh}^{-1}\left(\prod_{i' \in \partial j \backslash i}\mathrm{tanh}(M_{v_{i'} \rightarrow f_j}/2)\right)
\end{equation}

\noindent where $s_j$ is 0 if check $j$ is satisfied and 1 if it is unsatisfied, and $\partial j$ denotes the set of neighbours of factor node $j$. New variable-to-factor messages are then calculated as

\begin{equation}
    M_{v_i \rightarrow f_j} = l_i^0 + \sum_{j' \in \partial i \backslash j} M_{f_{j'} \rightarrow v_i}
\end{equation}

\noindent and updated log-likelihood ratios can be calculated after each iteration as 

\begin{equation}
    l_i = l_i^0 + \sum_{j \in \partial i} M_{f_j \rightarrow v_i}.
\end{equation}

If $l_i > 0$ then we expect that bit $i$ is error-free ($P(v_i = 0) > P(v_i = 1)$), and otherwise we expect it has been flipped. In global implementations of BP the syndrome from these expected errors is compared to the measured syndrome after each iteration and the algorithm terminates when these syndromes match (or if some specified iteration limit is reached without finding a valid correction, in which case failure is declared). We will instead consider a local version of BP where the algorithm always performs a specified number of iterations and then applies a correction to all bits where $l_i < 0$. When applying BP to quantum codes we lose almost nothing by doing this as we generally do not expect BP to find an exact correction (so it will usually run to the iteration limit anyway), and any slowdown coming from wasted iterations will likely be compensated for by the ability to perform all decoding calculations locally.  

BP can also be straightforwardly adapted to account for stabiliser measurement errors~\cite{li_numerical_2020}. This is done by adding a single new variable node for each factor node. These new nodes represent measurement errors rather than qubit errors and are connected only to their corresponding factor node and nothing else. In some codes the syndromes themselves can be interpreted as an error-correcting code that protects against measurement errors; for instance, in the 3DTC all valid $X$ error syndromes are closed loops of $Z$ stabilisers, so strings with open endpoints indicate the presence of measurement errors. This kind of structure can be incorporated into the factor graph through the use of ``meta-checks'', which enforce parity constraints on only the measurement error variable nodes. Despite the existence of such meta-checks in the 3DTC we do not include them in our simulations since they are not a general feature of quantum codes and we wish to assess whether BP works even in their absence. 

Our simulation results show that pure BP decoding does indeed work, and even retains much of the performance of BP+OSD. We once again use a phenomenological noise model with $p=q$, and after 500 code cycles with 30 BP iterations per code cycle we observe a threshold of $\sim 5.1\%$, beating both \pflip{} and the two-stage version of BP+OSD. As with our simulations of \flip{} and \pflip{}, decoding success/failure is decided via a majority vote on a complete disjoint set of logical $Z$ operators. 

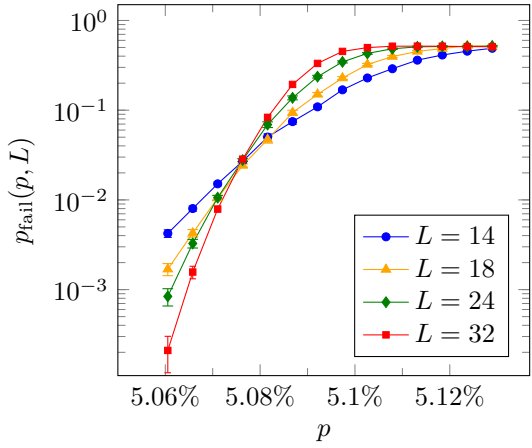
\begin{figure}
    \centering
        \begin{tikzpicture}
            \begin{axis}[
                xlabel={$p$},
                ylabel={$p_{\mathrm{fail}}(p,L)$},
                ymode=log,
                width=7cm,
                height=6.5cm,
                ymin=0.00011,
                ymax=1.5,
                xmin=0.0505,
                xtick={0.0506,0.0508,0.0510,0.0512},
                minor xtick={0.0507,0.0509,0.0511},
                xticklabels={$5.06\%$,$5.08\%$,$5.1\%$,$5.12\%$},
                legend style={
                    at={(0.95,0.05)},
                    anchor=south east
                }
            ]
            \addplot[
                color=blue,
                mark=*,
                mark size=1.5,
                error bars/.cd,
                y dir=both,
                y explicit
            ] table [x=p, y=pfail, y error=err, col sep=comma] {data/bp/processed14.csv};
            \addplot[
                color=orange,
                mark=triangle*,
                mark size=2,
                error bars/.cd,
                y dir=both,
                y explicit
            ] table [x=p, y=pfail, y error=err, col sep=comma] {data/bp/processed18.csv};
            \addplot[
                color=green,
                mark=diamond*,
                mark size=2,
                error bars/.cd,
                y dir=both,
                y explicit
            ] table [x=p, y=pfail, y error=err, col sep=comma] {data/bp/processed24.csv};
            \addplot[
                color=red,
                mark=square*,
                mark size=1.2,
                error bars/.cd,
                y dir=both,
                y explicit
            ] table [x=p, y=pfail, y error=err, col sep=comma] {data/bp/processed32.csv};
            \legend{$L=14$, $L=18$, $L=24$, $L=32$}
            \end{axis}
        \end{tikzpicture}
    \caption{Performance of BP (30 iterations per code cycle) in the 3DTC for physical $X$ error rate $p$ equal to $Z$ stabiliser measurement error rate $q$. $p_{fail}$ is the logical $X$ error rate calculated via a majority vote on the measurement outcomes of a complete set of disjoint logical $Z$ operators after 500 code cycles.}
    \label{fig:bp}
\end{figure}

An analysis of the 3DTC errors which cannot be decoded by BP was given in~\cite{higgott_improved_2022}. The most common are ``half-cube'' errors which are equivalent to half of an $X$ stabiliser generator such as \cref{subfig:uncorrec_a} and \cref{fig:uncorrec_dynamic}. In each of these cases the $Z$ syndrome loop divides the $X$ stabiliser into two equal halves and each of these halves is an equally likely candidate error for the observed syndrome. BP therefore cannot choose between the two halves and all qubits in the support of the $X$ stabiliser will have $l_i > 0$, meaning that no correction will be applied to any of these qubits and the error will be unchanged. Therefore, as with the low-weight uncorrectable errors we considered for \flip{}, BP can fail to correct these errors but cannot make them larger or otherwise make actively harmful mistakes. In later code cycles additional single qubit errors may occur in the vicinity of these half-cubes and break the symmetry between the two equivalent corrections, making the error correctable for BP. In order to grow the error while leaving it uncorrectable an additional (at least) two-qubit error must occur, transforming the original half-cube error (e.g. \cref{subfig:uncorrec_a}) into an error equivalent to half of a pair of cubes (e.g. \cref{subfig:newerrors_c}). 

We note also that, unlike \flip{}, BP does not struggle to correct large planar errors such as the one shown in \cref{subfig:uncorrec_b}. Additionally, even for relatively high-weight uncorrectable errors equivalent to half of a very large $X$ stabiliser, a single additional $X$ error is sufficient to break the symmetry between the two possible corrections and make the error correctable for BP (at least in principle). We therefore expect the threshold for BP to be considerably more sustainable than the one observed for \flip{} (although the question of whether or not this is a true ``sustainable threshold''\cite{brown_fault-tolerant_2016} would require additional simulations to answer). 

\subsection{p-BP}

\begin{figure}
    \centering
        \begin{tikzpicture}
            \begin{axis}[
                xlabel={$p$},
                ylabel={$p_{\mathrm{fail}}(p,L)$},
                ymode=log,
                width=7cm,
                height=6.5cm,
                ymin=0.0004,
                ymax=1.5,
                xmin=0.0540,
                xtick={0.0542,0.0546,0.055},
                minor xtick={0.0544,0.0548,0.0552},
                xticklabels={$5.42\%$,$5.46\%$,$5.5\%$,$5.54\%$},
                legend style={
                    at={(0.95,0.05)},
                    anchor=south east
                }
            ]
            \addplot[
                color=blue,
                mark=*,
                mark size=1.5,
                error bars/.cd,
                y dir=both,
                y explicit
            ] table [x=p, y=pfail, y error=err, col sep=comma] {data/p-bp/processed14.csv};
            \addplot[
                color=orange,
                mark=triangle*,
                mark size=2,
                error bars/.cd,
                y dir=both,
                y explicit
            ] table [x=p, y=pfail, y error=err, col sep=comma] {data/p-bp/processed18.csv};
            \addplot[
                color=green,
                mark=diamond*,
                mark size=2,
                error bars/.cd,
                y dir=both,
                y explicit
            ] table [x=p, y=pfail, y error=err, col sep=comma] {data/p-bp/processed24.csv};
            \addplot[
                color=red,
                mark=square*,
                mark size=1.2,
                error bars/.cd,
                y dir=both,
                y explicit
            ] table [x=p, y=pfail, y error=err, col sep=comma] {data/p-bp/processed32.csv};
            \legend{$L=14$, $L=18$, $L=24$, $L=32$}
            \end{axis}
        \end{tikzpicture}
    \caption{Performance of p-BP in the 3DTC for physical $X$ error rate $p$ equal to $Z$ stabiliser measurement error rate $q$. The decoder uses 20 iterations of BP followed by six iterations of \pflip{} (on the first and fourth iteration) and \flip{} (otherwise). $p_{fail}$ is the logical $X$ error rate calculated via a majority vote on the measurement outcomes of a complete set of disjoint logical $Z$ operators after 500 code cycles.}
    \label{fig:p-bp}
\end{figure}
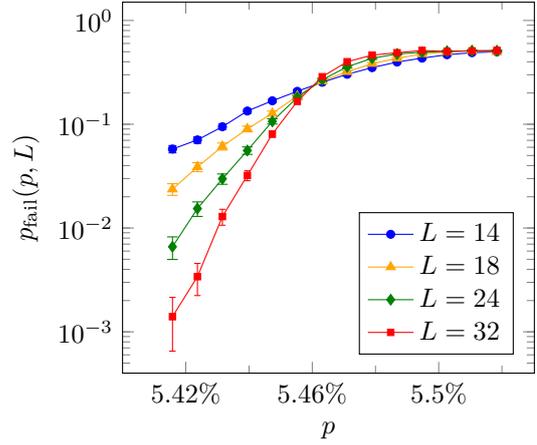

Finally, we consider a hybrid decoder that uses a combination of \flip{}, \pflip{} and BP. We refer to this decoder as p-BP. p-BP operates by first running BP to some specified number of iterations, applying corrections to any qubits with $l_i < 0$ and then running \flip{} and \pflip{} on some desired schedule. 

p-BP still cannot deterministically correct all constant-weight errors but it outperforms each of its component decoders in isolation as each component helps to cover the weaknesses of the others. \flip{} and \pflip{} complement each other in the ways discussed previously, while BP complements \flip{}/\pflip{} by correcting large planar errors and \pflip{} assists BP by probabilisitically correcting half-cube errors. We simulated the performance of p-BP using 20 iterations of BP and the following schedule for \flip{}/\pflip{}: \pflip{} $\rightarrow$ \flip$\times2$ $\rightarrow$ \pflip{} $\rightarrow$ \flip{}$\times2$. The results are shown in \cref{fig:p-bp}, where we observe a threshold of $\sim5.46\%$. Note also that we have used only 20 BP iterations in these simulations as opposed to the 30 iterations we used when decoding with BP alone. We observed that (unlike in the case of BP) reducing the iteration number from 30 to 20 did not appear to significantly affect p-BP's performance. The computational cost of running six rounds of \flip{}/\pflip{} is much lower than the cost of 10 additional rounds of BP, so p-BP outperforms BP not only in terms of accuracy but also in terms of runtime. 

We note that p-BP is analogous to the hybrid BP + {\tt small-set-flip} (SSF) decoder described in~\cite{grospellier_combining_2021}, but with SSF replaced by a combination of regular \flip{} and \pflip{}. However, in that work it was observed that in the presence of measurement noise the performance of BP+SSF was inferior to the performance of pure BP, whereas here the opposite is true. 

\section{Discussion}
\label{section:discussion}
It is commonly assumed in quantum error correction literature that a useful decoder must be able to correct all errors which have a size that is constant with respect to the code distance. In combination with the results for BP obtained in~\cite{grospellier_combining_2021}, our results indicate that this assumption may be unnecessarily strict and that competitive thresholds can be achieved even by decoders which fail to correct some very low weight errors. Additionally, simple modifications to these decoders can enable them to correct these low weight errors with finite probability, boosting performance without compromising locality or decoding time. We have studied only two examples (\flip{} and BP), but it is likely that there exist many similar decoders which were assumed to be unsuitable for quantum codes but which, in reality, might achieve practical thresholds. The threshold obtained by p-BP in the 3DTC is roughly 3/4 of the best known threshold for this code (using BP+OSD), but p-BP is fully local and so has a runtime that is constant in the code size, as opposed to BP+OSD which runs in time $O(n^3)$~\cite{higgott_improved_2022}. This is of great practical relevance as slow decoding algorithms can be a significant bottleneck when trying to improve the clock speed of a fault-tolerant quantum computer~\cite{terhal_quantum_2015,das_scalable_2020,chamberland_techniques_2022,skoric_parallel_2022}, and the severity of error propagation through transversal non-Clifford gates is also highly dependent on code cycle length and decoding time~\cite{scruby_non-pauli_2022}. We also remark that the threshold for p-BP is around twice as high as the threshold for the sweep decoder (another local decoder for the 3DTC) presented in~\cite{kubica_cellular-automaton_2019,vasmer_cellular_2021}, although it is possible that increasing the number of applications of the sweep rule per code cycle could raise those thresholds to be similar to the ones presented here. There are also a number of fairly simple improvements that could be made to p-BP that would likely increase its threshold further, such as using a longer and more optimised \flip{}/\pflip{} schedule and adding metacheck factors to BP. Performing readout via majority vote on logical $Z$ measurement outcomes is also suboptimal, and more accurate readout could be achieved by post-processing single-qubit measurement data with a decoder that can correct all constant-size errors. 

The most obvious open question regarding decoders of this type is whether or not they are able to produce sustainable thresholds. It may be the case that all such decoders result in slowly-growing error regions which will eventually cause the code to fail after many code cycles, but if such errors do exist then the timescale for their growth is too slow for them to be observed in feasible simulations. Instead these decoders should be studied analytically and, if possible, rigorous proofs for the existence of thresholds should be obtained. The recovery procedure for the 4DTC (which is essentially equivalent to \pflip{}) was studied analytically via comparison to a heat bath algorithm~\cite{dennis_topological_2002} and so this analysis may generalise to other decoders of this type. 

Another question is whether or not these kinds of decoding strategies can be successfully applied to other LDPC quantum codes. These more general codes have the potential to massively reduce the overhead associated with fault-tolerantly encoding large numbers of logical qubits and so fast and accurate decoders for them would be extremely valuable. SSF has been shown to work with quantum expander \cite{leverrier_quantum_2015} and hypergraph product codes \cite{grospellier_numerical_2019}, and many of the proposed decoders for the recently discovered \textit{good LDPC codes}~\cite{panteleev_asymptotically_2022,leverrier_quantum_2022} are also based on SSF~\cite{gu_efficient_2022,leverrier_decoding_2022,dinur_good_2022}. SSF is itself based on \flip{} and operates using the same principle of attempting to locally reduce syndrome weight, so we might also expect \flip{} and \pflip{} to be applicable to these codes. The runtime of SSF is exponential in the check degree of the code, whereas \flip{}/\pflip{} are linear in the bit degree and BP is quadratic in the bit and check degrees, so p-BP may be faster than SSF-based decoders in practise (even though all of these decoders are linear in terms of the blocklength of the code). Simulations studying the performance of p-BP in this setting will be the topic of future work.

It is our hope that the results presented here will lead to the discovery of new, more practical decoders which will reduce the demanding requirements placed on quantum hardware and classical control software by current decoding algorithms. 

\section*{Acknowledgements}
The authors acknowledge support from the JST Moonshot R\&D Grant [grant number JPMJMS2061]. Numerical results presented in this work were obtained using the HPC resources provided by the Scientific Computing and Data Analysis section of the Research Support Division at OIST. TRS is grateful to M. Vasmer and J. Roffe for comments on the manuscript, and also acknowledges valuable discussions with O. Higgott, N. P. Breuckmann, M. Vasmer, J. Roffe and A. Krishna. Some of these discussions occurred during a visit by TRS to Perimeter Institute that was partially funded by Perimeter Institute. Research at Perimeter Institute is supported in part by the Government of Canada through the Department of Innovation, Science and Economic Development Canada and by the Province of Ontario through the Ministry of Colleges and Universities. 

\bibliographystyle{unsrtnat}
\bibliography{references}

\newpage

\section*{Appendices}

\appendix

\section{Extra Simulation Details}
\label{appendix:sim}
All code used in obtaining the results presented in this paper can be found at \cite{source_code}

\subsection{GPU Implementation and Performance}

As discussed in the main text, one of the main strengths of the decoders we had studied is their parallelisability. This allows not only for dramatic improvements in decoding time in a physical fault-tolerant quantum computer, but also for much faster numerical simulations of decoding performance. In order to take advantage of this all simulations carried out as part of this work were written using CUDA, a GPU programming language that allows for the full parallelisation of the simulated decoders. Using an HPC system with access to a large number of GPUs we are then able to obtain the data for plots such as \cref{fig:flip_1-2} in a few hours, and for plots such as \cref{fig:bp} and \cref{fig:p-bp} in about a day or so, despite the large size of the codes involved (the longer runtime required for the latter two is due to the fact that, although all studied decoders are asymptotically constant time, a single BP iteration is more computationally expensive than a single iteration of \flip{}).

In principle, the runtime of these simulations should be independent of the code size and should depend only on the number of simulated decoding cycles. In practise we are limited by the number of GPUs available, and when the number of qubits in the code becomes greater than the number of available GPU threads some of the decoding has to be serialised. For this reason the simulations of the $L=32$ codes are noticably slower than those for the smaller code sizes, and so we do not simulate any codes larger than this. 

\subsection{Logical Error Correlations}

The plots presented in the main text show logical error rates for only one of the three qubits encoded by the 3D toric code. Logical error rates for the other qubits will be identical as, for a given $L$, the distance is the same for all three encoded qubits and the error model and decoder are symmetric along all three axes of the code. However, correlations between these three encoded qubits can still occur. For instance, consider the uncorrectable error shown in \cref{subfig:uncorrec_a}, which is equivalent to three qubits of an $X$ stabiliser and is supported on three different faces of a cube. Because each of these faces lies in a different plane, each one can contribute to the support of a different logical $X$ operator and so we expect the probability of a logical error affecting all three encoded qubits to be increased, while the probability of errors affecting only a single encoded qubit will be reduced. In \cref{table:correlations} we show the frequency of occurrence of different logical $X$ error configurations after 500 decoding cycles with p-BP when $p=0.544$ (so very slightly below the numerically observed threshold), and we can see that these data confirm our expectations. 

\begin{table}[]
    \centering
    \begin{tabular}{c|c}
        Error Configuration & \# Occurrences \\
        \hline
        000 & 731 \\
        100 & 52 \\
        010 & 53 \\
        001 & 62 \\
        110 & 21 \\
        101 & 30 \\
        011 & 26 \\
        111 & 25
    \end{tabular}
    \caption{Frequency of occurrence of each possible configuration of logical $X$ errors on the three qubits encoded in an $L=14$ 3D toric code after 500 code cycles of decoding with p-BP and physical error rate (= measurement error rate) $p=0.544$. 1000 simulations were performed in total. The frequency of multi-qubit logical errors is higher than would be expected if each of the three encoded qubits experienced logical errors independently.}
    \label{table:correlations}
\end{table}

\end{document}